\documentstyle[floats,preprint,aps,epsf,amssymb]{revtex}
\begin{document}

\title{Evaporation of a two-dimensional charged black hole}

\author{Amos Ori}
\address {Department of Physics,
          Technion---Israel Institute of Technology, Haifa, 32000, Israel}
\date{\today}
\maketitle

\begin{abstract}
We construct a dilatonic two-dimensional model of a charged black hole. The
classical solution is a static charged black hole, characterized by two
parameters, $m$ and $q$, representing the black hole's mass and charge.
Then
we study the semiclassical effects, and calculate the evaporation rate of
both $m$ and $q$, as a function of these two quantities. Analyzing this
dynamical system, we find two qualitatively different regimes, depending on
the electromagnetic coupling constant $g_{A}$. If the latter is greater than
a certain critical value, the charge-to-mass ratio decays to zero upon
evaporation. On the other hand, for $g_{A}$ smaller than the critical value,
the charge-to-mass ratio approaches a non-zero constant that depends on
$g_{A}$ but not on the initial values of $m$ and $q$.
\end{abstract}

\section{Introduction}

Dilatonic two-dimensional models are very useful in studying various aspects
of black holes (BHs). Callan, Giddings, Harvey and Strominger (CGHS) \cite
{CGHS} introduced a dilatonic two-dimensional model to investigate the
evaporation of a BH. They further used the two-dimensional model to explore
the final stage of the evaporation process. The latter, more ambitious, goal
of clarifying the endpoint of evaporation has proved to be illusive and hard
to achieve\cite{russo}. Nevertheless, the two-dimensional model provided a
useful description of the evaporation process in its initial phase, i.e., as
long as the BH's mass is sufficiently large. Since then, many authors
studied various aspects and various variants of CGHS's model.

The two-dimensional model studied by CGHS was that of an uncharged BH. 
The
main objective of this paper is to develop an analogous model to describe
the evaporation of an {\it electrically charged} two-dimensional BH. Our
main motivation for studying charged BHs stems from the basic features of
the classical, macroscopic, 4-dimensional BH solutions. The spacelike
singularity of the Schwarzschild geometry disappears when an electric
charge is added to the BH, and an inner horizon (IH) forms instead.
Remarkably, the same situation occurs when an angular momentum is added 
to
the BH (instead of an electric charge): In the Kerr solution, too, there
is an inner horizon and no spacelike singularity. This observation is very
relevant to reality, because realistic astrophysical BHs are believed to
be rapidly spinning \cite {bardeen}. Spherically symmetric charged BHs
thus provide a useful toy model for exploring various aspects related to
the inner structure of the more realistic, spinning BHs (see e.g. Ref.
\cite{PI}).

In the last decade several authors investigated two-dimensional models of
charged BHs (see e.g. \cite{MNY},\cite{NO},\cite{Elizalde},\cite{nojiri.new}
). McGuigan, Nappi, and Yost \cite{MNY} studied such a classical model, with
a dilaton coupling to the electric field. They considered a coupling ${\rm e}
^{-2\phi }$ of the dilaton field $\phi $ to the electromagnetic term in the
action (see below). They found that, just as in the four-dimensional case,
charged BHs admit an inner horizon instead of a spacelike singularity.
However, their model does not include semiclassical effects, which are
necessary for describing the BH's evaporation.

Later, Nojiri and Oda (NO) \cite{NO} considered a slightly modified
dilatonic model, in which the dilaton coupling to the electric field is $
{\rm e}^{2\phi }$. In their model there is a large number $N$ of chiral
fermion fields (instead of the $N$ scalar matter fields considered in Ref.
\cite{CGHS}), which couple to both the curvature and the electromagnetic
field. NO first studied the structure of the classical BH solution. Then,
generalizing the method used by CGHS to the charged case, they added two
effective semiclassical correction terms to the action, and derived from
them the field equations at the semiclassical level. They have been able to
solve some of the semiclassical field equations, which allowed them to
analyze various aspects of the resultant semiclassical charged BH solutions.

Nojiri and Oda considered the dilaton coupling ${\rm e}^{2\phi }$ rather
than ${\rm e}^{-2\phi }$, primarily because the former coupling makes some
of the equations easier to solve. As it turns out, however, there is a
remarkable difference between the two models, already at the classical
level. The charged BHs with the coupling ${\rm e}^{2\phi }$ do {\it not}
admit an inner horizon \cite{NO} (instead they usually admit a spacelike
singularity, just as in the uncharged case). Since our main motivation for
considering charged BHs is to mimic the inner structure of the
four-dimensional spinning BHs, we find it important to elaborate on the
charged model with the dilaton coupling ${\rm e}^{-2\phi }$. (Note also that
${\rm e}^{-2\phi }$ is the coupling which emerges as the effective action in
the low-energy limit of string theory.\cite{NO})

Motivated by these considerations, in this paper we start from the model
developed by NO, and modify the dilaton coupling to the electromagnetic
field from ${\rm e}^{2\phi }$ to ${\rm e}^{-2\phi }$. The resultant field
equations are harder to solve at the semiclassical level. Nevertheless, it
is possible to solve the equations describing the semiclassical effects in
the adiabatic approximation, i.e. in the approximation where the background
geometry is described by the static, non-evaporating, BH. \cite{adiabatic} 
This approximation
appears to be valid as long as the black hole is macroscopic, in which case
the evaporation rate is small (namely, the relative change in the mass or
charge during a dynamical time scale is $<<1)$. It is this macroscopic
domain which will concern us throughout this paper. Thus, on the background
of a classical static BH solution (with given mass and charge), we solve the
semiclassical equations and derive from them the semiclassical fluxes of
both energy-momentum and charge. This allows us to determine the
evaporation
rate of both the mass and charge of the BH.

We start in section II by writing the classical action and the corresponding
field equations. We define new variables, which are used to simplify the
equations and to present their general solution. This general solution is a
two-parameter family of static black hole solutions, uniquely characterized
by the two parameters $m$ and $q$, which represent the BH's mass and
charge.
This classical solution was already given in Ref. \cite{MNY}, though in
different coordinates. Here we construct the classical solution in
double-null coordinates, which are more suitable for the subsequent
semiclassical calculations. We also extend the classical solution to include
an outgoing (or ingoing) null fluid. (This extension is useful for
describing the geometry of the evaporating BH at large distance from the
horizon.)

In section III we consider the semiclassical effects. Following NO, we
analyze the semiclassical effects by adding two effective correction terms
to the classical action, expressed in terms of two new variables $Z$ and $Y$
(these variables describe the semiclassical fluxes of energy-momentum and
charge, respectively). From this action we derive the semiclassical field
equations. Then we solve the field equations for the two quantum variables
$
Z $ and $Y$, assuming a background geometry of a static, classical BH. This
solution yields an explicit expression for the fluxes of energy-momentum and
charge, at any location (both outside and inside the BH), as a function of $
m $ and $q$. Based on these fluxes, in section IV we calculate the rate of
evaporation of the mass and charge. We obtain a closed system of two
first-order equations, describing $\dot{m}$ and $\dot{q}$ as functions of
$m$
and $q$, where an overdot denotes a derivative with respect to the external
time. Both $\dot{m}$ and $\dot{q}$ are found to be negative, as one should
expect. We then analyze this dynamical system, and find two qualitatively
different regimes. If the electromagnetic coupling constant $g_{A}$ is
larger than a certain critical value, the charge-to-mass ratio decays to
zero upon evaporation (as a certain power of the mass, which depends on $
g_{A}$). On the other hand, for $g_{A}$ smaller than the critical value, the
charge-to-mass ratio approaches a non-zero constant, that depends on
$g_{A}$
but not on the initial values of $m$ and $q$.

In section V we summarize and discuss our results. It should be emphasized
that no attempt is made in this paper to investigated the final stage of
evaporation. The goal here is to study the semiclassical evolution of the BH
in the {\it macroscopic} phase, i.e. as long as the BH's mass is much larger
than a certain mass. It is this phase in which the above mentioned adiabatic
approximation is valid. In section V we further discuss this validity
criterion for the adiabatic approximation, and find the range of mass values
for which this approximation can be used.

\section{Classical solutions}

\subsection{field equations}

We start with the classical action $S_{c}$ given by NO \cite{NO}, which
includes an electromagnetic field coupled to charged matter represented by
$
N $ left-handed chiral fermions:
\begin{eqnarray}
S_{c} &=&{\frac{1}{2\pi }}\int d^{2}x\,\sqrt{-g}\{{\rm e}^{-2\phi
}(R+4(\nabla \phi )^{2}+4\lambda ^{2})  \nonumber \\
&&-{\frac{{\rm e}^{a\phi }}{g_{A}^{2}}}F^{2}-\sum_{j=1}^{N}i\bar{\Psi}
_{j}\gamma ^{\mu }(D_{\mu }-iA_{\mu })\Psi _{j}\}\ .  \label{eq101}
\end{eqnarray}
Here $\phi $ is a dilaton field, $\Psi_j={\psi_j \choose 0}$ are the $
N$ left-handed chiral fermion fields, $g_{A}$ is the electromagnetic
coupling constant, and $D_{\mu }$ denotes a covariant derivative. The
Maxwell tensor $F_{\mu \nu }$ is given by $F_{\mu \nu }=\partial _{\mu
}A_{\nu }-\partial _{\nu }A_{\mu }$, with $F^{2}\equiv F_{\mu \nu }F^{\mu
\nu }$. The coupling of the dilaton to the electromagnetic field term is
written here, in a quite general form, as ${\rm e}^{a\phi }$. In Ref. \cite
{NO}, NO only analyzed the case $a=2$. Here, for the reasons explained in
the Introduction, we shall consider the case $a=-2$.

Following NO, we use the light-cone gauge for the electromagnetic field,
namely
\begin{equation}
A_{u}=0\ .  \label{eiib}
\end{equation}
We also use double-null coordinates $u,v$, with
\begin{equation}
\,g_{uv}=-{\frac{1}{2}}{\rm e}^{2\rho }\,~,~g_{uu}=g_{vv}=0.
\label{eq102}
\end{equation}
The action (\ref{eq101}) (with $a=-2$) then reduces to
\begin{eqnarray}
S_{c} &=&{\frac{1}{2\pi }}\int d^{2}x\,\{{\rm e}^{-2\phi }(4\rho
_{,uv}-8\phi _{,u}\phi _{,v}+2\lambda ^{2}{\rm e}^{2\rho })  \nonumber \\
&&+{\frac{4}{g_{A}^{2}}}{\rm e}^{-2(\phi +\rho )}F_{uv}^{2}+{\frac{i}{2}}
\sum_{j=1}^{N}\psi _{j}^{*}\partial _{v}\psi _{j}\}\ ,  \label{eiii}
\end{eqnarray}
where $F_{uv}=\partial _{u}A_{v}$. The Einstein equations are given by
\begin{eqnarray}
0 &=&T_{vv}={\rm e}^{-2\phi }(4\rho _{,v}\phi _{,v}-2\phi
_{,vv})+{\frac{i}{4
}}\sum_{j=1}^{N}(\psi _{j}^{*}\partial _{v}\psi _{j}-\partial _{v}\psi
_{j}^{*}\psi _{j})  \nonumber \\
&&+{\frac{1}{2}}A_{v}\sum_{j=1}^{N}\psi _{j}^{*}\psi _{j}\ ,  \label{eva} \\
0 &=&T_{uu}={\rm e}^{-2\phi }(4\rho _{,u}\phi _{,u}-2\phi _{,uu})\ ,
\label{evb}
\end{eqnarray}
\begin{equation}
0=T_{uv}={\rm e}^{-2\phi }(2\phi _{,uv}-4\phi _{,u}\phi _{,v}-\lambda ^{2}
{\rm e}^{2\rho })+{\frac{2}{g_{A}^{2}}}{\rm e}^{-2(\phi +\rho
)}F_{uv}^{2}\ .
\label{evi}
\end{equation}
The dilaton equation of motion is
\begin{equation}
0=-4\phi _{,uv}+4\phi _{,u}\phi _{,v}+2\rho _{,uv}+\lambda ^{2}{\rm e}
^{2\rho }+{\frac{2}{g_{A}^{2}}}{\rm e}^{-2\rho }F_{uv}^{2}\ ,  \label{eviii}
\end{equation}
the Maxwell equations are
\begin{eqnarray}
0 &=&{\frac{8}{g_{A}^{2}}}\partial _{v}({\rm e}^{-2(\phi +\rho )}F_{uv})+{
\frac{1}{2}}\sum_{j=1}^{N}\psi _{j}^{*}\psi _{j}\ ,  \nonumber \\
0 &=&{\frac{8}{g_{A}^{2}}}\partial _{u}({\rm e}^{-2(\phi +\rho )}F_{uv})\ ,
\label{evii}
\end{eqnarray}
and the fermion fields satisfy

\begin{equation}
0=\partial _{u}\psi _{j}\ .  \label{fermion}
\end{equation}

We shall consider here solutions free of any classical matter. \footnote{
Presumably the charged BH was created by the collapse of the fermionic
matter. Here, however, we are interested in the evaporation of the BH and
not in its creation.} That is, we shall only consider here the trivial
solution

\begin{equation}
\psi _{j}=0  \label{nomatter}
\end{equation}
to Eq. (\ref{fermion}). The Maxwell equations then reduce to
\begin{equation}
0=\partial _{u}({\rm e}^{-2(\phi +\rho )}F_{uv})=\partial _{v}({\rm e}
^{-2(\phi +\rho )}F_{uv})\ ,
\end{equation}
namely,
\begin{equation}
(\sqrt{2}/g_{A})~{\rm e}^{-2(\phi +\rho )}F_{uv}={\rm const}\equiv
\lambda q.
\label{Q}
\end{equation}
Substituting Eqs. (\ref{fermion}) and (\ref{Q}) in the above
system(\ref{eva}
-\ref{eviii}), we obtain a simpler closed system, which includes the three
Einstein equations

\begin{equation}
0=T_{uv}={\rm e}^{-2\phi }(2\phi _{,uv}-4\phi _{,u}\phi _{,v}-\lambda ^{2}
{\rm e}^{2\rho }+\lambda ^{2}{q}^{2}{\rm e}^{4\phi +2\rho })\,,
\label{tuv}
\end{equation}
\begin{eqnarray}
0 &=&T_{vv}={\rm e}^{-2\phi }(4\rho _{,v}\phi _{,v}-2\phi _{,vv})\,,
\label{tvv} \\
0 &=&T_{uu}={\rm e}^{-2\phi }(4\rho _{,u}\phi _{,u}-2\phi _{,uu})\ ,
\label{tuu}
\end{eqnarray}
and the dilaton equation
\begin{equation}
0=-4\phi _{,uv}+4\phi _{,u}\phi _{,v}+2\rho _{,uv}+\lambda ^{2}{\rm e}
^{2\rho }+\lambda ^{2}{q}^{2}{\rm e}^{4\phi +2\rho }\,.  \label{eq-fi}
\end{equation}
Note that the two equations (\ref{tuv}) and (\ref{eq-fi}) -- to which we
shall refer as the {\it evolution equations} -- are hyperbolic, and are
hence sufficient for determining the evolution of the two unknowns $\rho $
and $\phi $ from prescribed initial data. [The other two equations (\ref{tvv}
,\ref{tuu}) -- the {\it \ constraint equation} -- are consistent with the
evolution equations: Any solution of the evolution equations whose initial
data are consistent with the constraint equations, will also satisfy the
constraints in the entire domain of dependence.]

\subsection{New variables}

To further simplify the analysis, we define the new variables

\begin{equation}
\ R={\rm e}^{-2\phi }~,~S=2(\rho -\phi ).  \label{RS}
\end{equation}
The evolution equations then reduce to

\begin{equation}
R_{,uv}=\lambda ^{2}({q}^{2}/R^{2}-1)~{\rm e}^{S},  \label{eqR}
\end{equation}

\begin{equation}
\ S_{,uv}=(-2\lambda ^{2}{q}^{2}/R^{3})~{\rm e}^{S}.  \label{eqS}
\end{equation}
The two constraint equations also take a simple form:

\begin{eqnarray}
\ 0 &=&T_{uu}=R_{,uu}-R_{,u}S_{,u}~,  \label{Ruu} \\
0 &=&T_{vv}=R_{,vv}-R_{,v}S_{,v}~.  \label{Rvv}
\end{eqnarray}

The two evolution equations (\ref{eqR},\ref{eqS}) can be expressed in a
compact form as
\begin{equation}
R_{,uv}=F(R)~{\rm e}^{S}\quad ,\quad S_{,uv}=F_{,R}~{\rm e}^{S}\,.
\label{universal}
\end{equation}
where
\begin{equation}
F(R)=\lambda ^{2}({q}^{2}/R^{2}-1)\,.  \label{F(R)}
\end{equation}
As it turns out, the system (\ref{universal}) is rather universal, as
various general-relativistic models (e.g. several two-dimensional BH models,
the three-dimensional BTZ model, and various spherically-symmetric
four-dimensional models) satisfy the same form of hyperbolic system, with
each model having its own function $F(R)$. For example, in the
spherically-symmetric four-dimensional model of a charged BH with (or
without) a cosmological constant, if one defines $R\equiv r^{2}$ and ${\rm
e}
^{S}\equiv rg_{uv}$ (where $r$ is the area coordinate and $u,v$ are two null
coordinates), the two Einstein evolution equations take exactly the form (
\ref{universal}) with
\begin{equation}
F(R)=aR^{1/2}+bR^{-1/2}+cR^{-3/2}\,\qquad \rm{(four-dimensional)}\,,
\label{4D}
\end{equation}
where $a,b,c$ are constants ($a$ and $c$ represent the contributions of the
cosmological constant and charge, respectively).

The non-linear hyperbolic system (\ref{universal}) [for a rather generic
function $F(R)$] may serve as a useful mathematical toy model for studying
various aspects of the theory of black holes, like gravitational collapse,
singularity formation, and the no-hair principle. This, however, is beyond
the scope of the present paper. Here we shall merely use the $R,S$ variables
to simplify the analysis, as the system (\ref{universal}) does not include
first-order derivatives. We shall also use a few general features of this
system -- e.g. the form of its static black-hole solutions.

The generic solution of Eq. (\ref{universal}) does not necessarily satisfy
the constraint equations (\ref{Ruu},\ref{Rvv}). Of course, we are primarily
interested here in the subclass of solutions which do satisfy the constraint
equations, to which we shall refer as the {\it vacuum-like} solutions. Apart
from the gauge freedom (i.e. the freedom to re-parametrize each of the two
null coordinates $u,v$), this subclass is a one-parameter family of
solutions [for a given $F(R)$ ], parametrized by the mass. These vacuum-like
solutions are, in fact, the static black-hole solutions of the model. (For
example, in the four-dimensional spherical electrovacuum case, these are the
RN-deSittre family of solutions.) In the context of the specific model
considered in this paper [with $F(R)$ of Eq. (\ref{F(R)})], the vacuum-like
solutions are nothing but the two-dimensional static electrovacuum
solutions. We shall construct these static solutions, in double-null
coordinates, in the next subsection.

One may also be interested in the wider class of solutions to the evolution
equations (\ref{universal}), which do not necessarily satisfy the constraint
equations (\ref{Ruu},\ref{Rvv}). Such solutions may be interpreted as
spacetimes perturbed by ingoing and/or outgoing null fluids, leading to
non-vanishing contributions to $T_{vv}$ and/or $T_{uu}$. We shall name such
solutions as {\it radiative solutions}. For example, in the context of
spherically-symmetric, four-dimensional, charged BHs, the mass-inflation
solutions introduced in Ref. \cite{PI} belong to this class of radiative
solutions [with the function $F(r)$ of Eq. (\ref{4D})]. Note that although a
radiative solution does not satisfy all the vacuum field equations, its
evolution from Cauchy or characteristic initial data is completely
determined from the (vacuum!) evolution equations, which form a closed
hyperbolic system.

An important subclass of the radiative solutions are those which satisfy
{\it one} of the constraint equations, but not the other one. Such solutions
may be interpreted as spacetimes with a null fluid flowing in either the
outgoing or ingoing direction. We shall refer to these solutions as the {\it
Vaidya-like} solutions. For example, the geometry in the weak-filed region
(i.e. at large $R$) of an evaporating BH can be well approximated by an
outgoing Vaidya-like solution. In Appendix A we describe the construction of
the general Vaidya-like solution in double-null coordinates.

\subsection{The static black-hole solution}

The general solution of the above system (\ref{eqR} - \ref{Rvv}) (i.e. both
the evolution and constraint equations) is a family of two-dimensional
static, RN-like, black-hole solutions uniquely characterized by their mass
and charge. This general solution was presented in Ref. \cite{MNY} using
Schwarzschild-like coordinates. For the analysis below we shall need the
solution in double-null coordinates. We shall first describe the
construction of this solution for a general function $F(R)$, and then
restrict attention to our specific model, i.e. $F(R)=\lambda ^{2}({q}
^{2}/R^{2}-1)$.

For a general function $F(R)$, we define
\begin{equation}
H(R)\equiv -\int^{R}F(R^{\prime })dR^{\prime }\,.  \label{genh}
\end{equation}
The static, vacuum-like, solution only depends on the spatial coordinate,
which we denote $x$. We choose an Eddington-like gauge, such that $x=v-u$.
The solution is then given by

\begin{equation}
\ {\rm e}^{S}=H(R)~,~R_{,x}=H(R)~,~x=v-u\,\quad \quad (H>0).
\label{external}
\end{equation}
From the above definitions of $R$ and $S$, the metric function $g_{uv}$ is
given by

\begin{equation}
-2g_{uv}=e^{2\rho }=H/R\,.  \label{guvg}
\end{equation}
Note that this Eddington-like solution is only valid in the region outside
the BH where $H(R)>0$ -- this is the region which will primarily concern us
in this paper. The solution exhibits a coordinate singularity whenever $H$
vanishes [where $g_{uv}$ vanishes, and so does $\det (g)$ ]. The lines $H=0$
correspond to the horizons of the BH. These include the event horizon (EH),
and [for functions $H(R)$ which admit more than one zero] also the inner
horizon and/or the cosmological horizon. In the region inside the BH where $
H(R)$ is negative, the solution can be expressed in a very similar form --
see section III. Note that the solution includes one free parameter -- the
integration constant in Eq. (\ref{genh}) -- which is related to the BH's
mass (see below).

In the specific model considered in this paper, for which $F(R)=\lambda
^{2}(
{q}^{2}/R^{2}-1)$, we write $H(R)$ in the form

\begin{equation}
\ H(R)=\lambda ^{2}(R-2m+{q}^{2}/R)\,  \label{H}
\end{equation}
(for notational convenience we take here the integration constant to be $
-2m\lambda ^{2}$). The vacuum-like solution is thus uniquely determined by
the two parameters $m$ and $q$, which are related to the black-hole's mass
and charge, respectively. The root structure of the function $H(R)$ depends
on the ratio between $q$ and $m$. In this paper we shall consider
non-extreme charged black-hole solutions, i.e. solutions with $m>q>0$ (the
restriction to positive rather than negative $q$ does not cause any loss of
generality). The equation $H(R)=0$ then has two roots, at $R_{\pm }=m\pm
\sqrt{m^{2}-q^{2}}$, where $R_{+}$ corresponds to the EH and $R_{-}$
corresponds to the IH. The function $H(R)$ is positive outside the BH, i.e.
at $R>R_{+}$ (and also at $R<R_{-}$, but this range will not concern us in
this paper), and negative between the two horizons. As was mentioned
above,
the solution (\ref{external}) is only valid outside the BH, and a similar
one, valid inside the BH, is given in section III.

The surface gravity $\kappa _{+}$ of the EH is defined by
\begin{equation}
\kappa _{+}\equiv \frac{1}{2}(H_{,R})_{R_{+}}=\lambda ^{2}(1-{q}
^{2}/R_{+}^{2})/2\,.  \label{e73}
\end{equation}
For later convenience, we also express $\kappa _{+}$ in other useful forms:

\begin{equation}
\kappa _{+}=\lambda ^{2}(1-m/R_{+})=\lambda ^{2}\left[
(m^{2}-q^{2})^{1/2}/R_{+}\right] \,.  \label{e73b}
\end{equation}
It is remarkable that in the two-dimensional case, unlike the situation in
four-dimensional BHs, $\kappa _{+}$ only depends on $q/m$, and not on the
BH's size. One explicitly finds
\begin{equation}
\kappa _{+}=\lambda ^{2}\left[ 1-\left( 1+\sqrt{1-(q/m)^{2}}\right)
^{-1}\right] \,.  \label{e73a}
\end{equation}
Note that $\kappa _{+}$ is a decreasing function of $q/m$, and for $\,0\leq
q/m\leq 1$ it takes the values $\,0\leq \kappa _{+}\leq \lambda ^{2}/2$.

From Eq. (\ref{guvg}), the metric function $g_{uv}$ is given by

\begin{equation}
-2g_{uv}=\lambda ^{2}(1-2m/R+{q}^{2}/R^{2})\,.  \label{guv}
\end{equation}
At large $R$, this becomes $-2g_{uv}=\lambda ^{2}$. It is useful to
introduce new null coordinates

\begin{equation}
\tilde{u}=\lambda u,\;\tilde{v}=\lambda v\,,  \label{newuv}
\end{equation}
such that
\begin{equation}
-2g_{\tilde{u}\tilde{v}}=1-2m/R+{q}^{2}/R^{2}\,,  \label{gmod}
\end{equation}
which yields the desired asymptotic behavior, $-2g_{\tilde{u}\tilde{v}}=1$,
at large $R$ (this is also the type of gauge used by CGHS in the uncharged
case). We shall refer to the $\tilde{u},\tilde{v}$ coordinates as the {\it
asymptotically-flat double-null coordinates}.

Since $R\equiv {\rm e}^{-2\phi }$ is dimensionless, the two parameters $m$
and $q$ are dimensionless too. These two parameters are proportional to the
physical mass and charge of the BH, and we may refer to them as the
dimensionless mass and charge, respectively. [One of the motivations for
this association is the similarity of the metric function $g_{\tilde{u}
\tilde{v}}$ in Eq. (\ref{gmod}) to its counterpart in the standard
four-dimensional Reissner-Nordstrom solution.] Throughout this paper we
shall often refer to $m$ and $q$ simply as the BH's mass and charge (with
some abuse of the standard terminology). To avoid confusion, we shall
denote
the dimensionful, physical, mass and charge of the BH by $M$ and $Q$,
respectively. The physical mass $M$, which is the total energy content of
the system, is encoded in the asymptotic behavior of the metric tensor at
large distance. Since this large distance corresponds to the limit $
R\rightarrow \infty $, the physical mass will not be affected by the term $
q^{2}/R^{2}$ in the metric (\ref{gmod}). Hence $M$ must be a function of
$m$
and $\lambda $ solely. We can deduce $M(m,\lambda )$ from the form of the
mass parameter $M$ in the uncharged case, studied by CGHS. Comparing the
static solutions of the two models, one finds \footnote{
To relate $M$ in the CHGS model to our notation, we can compare the value
of
the dilaton field at the EH. In our notation we find at the EH (for $q=0$ ) $
{\rm e}^{-2\phi }\equiv R=2m$, whereas in CGHS we find at the EH ${\rm \
e}
^{-2\phi }=M/\lambda $; cf. Eq. (11) therein.}

\begin{equation}
M=2\lambda m\,.  \label{mass}
\end{equation}
The relation between $q$ and $Q$ may be revealed by comparing our static
solution to that given (in different coordinates) in Ref. \cite{MNY}. This
comparison shows that $Q$ is proportional to $\lambda q$ , in analogy with
Eq. (\ref{mass}). We have not clarified the constant relating $Q$ to $
\lambda q$ (however, the explicit expression for $Q$ is not required for the
analysis below).

\section{Semiclassical corrections}

\subsection{Semiclassical field equations}

In order to study the evaporation of the BH, we must consider the
semiclassical contribution to the energy-momentum tensor and to the
electromagnetic current. CGHS \cite{CGHS} showed that the semiclassical
contributions can be treated by adding an effective correction term to the
classical action. Nojiri and Oda \cite{NO} extended this method to the
charged case (they also modified the action by considering fermionic rather
than scalar matter fields). They found that the semiclassical contributions
coming from the conformal anomaly can be represented by adding two
correction terms to the classical action $S_{c}$:
\begin{equation}
S=S_{c}+S_{\rho }+S_{\chi }\ ,  \label{eq103}
\end{equation}
where

\begin{equation}
S_{\rho }={\frac{1}{2\pi }}\int d^{2}x\,\sqrt{-g}\{-{\frac{1}{2}}(\nabla
Z)^{2}+\sqrt{{\frac{N}{48}}}Z\,R\}\   \label{eq103b}
\end{equation}
and
\begin{equation}
S_{\chi }={\frac{1}{2\pi }}\int d^{2}x\,\{-{\frac{1}{2}}\sqrt{-g}(\nabla
Y)^{2}+\sqrt{{\frac{N}{2}}}Y\epsilon ^{\mu \nu }F_{\mu \nu }\}\ .
\label{eq103c}
\end{equation}
The term $S_{\rho }$, which emerges from the trace anomaly, contributes
to
the energy-momentum tensor. The second term $S_{\chi }$ comes from the
chiral anomaly, and it contributes to both the electromagnetic current and
the energy-momentum. Note the presence of two new variables, $Z$ and $Y$,
in
the correction terms. These two degrees of freedom were introduced by NO
in
order to allow the representation of the semiclassical effects by local
correction terms\cite{NO}.

Expressing the metric in the double-null form (\ref{eq102}) and using the
electromagnetic gauge (\ref{eiib}), the modified Einstein equations become

\begin{eqnarray}
0 &=&T_{vv}={\rm e}^{-2\phi }(4\rho _{,v}\phi _{,v}-2\phi _{,vv})+\hat{T}
_{vv}\,,  \label{svv} \\
0 &=&T_{uu}={\rm e}^{-2\phi }(4\rho _{,u}\phi _{,u}-2\phi _{,uu})+\hat{T}
_{uu}\,,  \label{suu}
\end{eqnarray}
\begin{equation}
0=T_{uv}={\rm e}^{-2\phi }(2\phi _{,uv}-4\phi _{,u}\phi _{,v}-\lambda ^{2}
{\rm e}^{2\rho }+{q}^{2}{\rm e}^{4\phi +2\rho })+\hat{T}_{uv}\ .
\label{exxvii}
\end{equation}
Here, $\hat{T}_{uu}$, $\hat{T}_{vv}$ , and $\hat{T}_{uv}$ represent the
semiclassical contributions to $T_{uu}$, $T_{vv}$, and $T_{uv}$,
respectively, which are given by

\begin{eqnarray}
\hat{T}_{vv} &=&{\frac{1}{2}}Z_{,v}^{\,2}-\sqrt{{\frac{N}{12}}}\rho
_{,v}Z_{,v}+{\frac{1}{2}}\sqrt{{\frac{N}{12}}}Z_{,vv}+{\frac{1}{2}}
Y_{,v}^{\,2}\,, \\
\hat{T}_{uu} &=&{\frac{1}{2}}Z_{,u}^{\,2}-\sqrt{{\frac{N}{12}}}\rho
_{,u}Z_{,u}+{\frac{1}{2}}\sqrt{{\frac{N}{12}}}Z_{,uu}+{\frac{1}{2}}
Y_{,u}^{\,2}\,, \\
\hat{T}_{uv} &=&-\sqrt{{\frac{N}{48}}}Z_{,uv}\ .
\end{eqnarray}
The Maxwell equations also get semiclassical source terms:
\begin{eqnarray}
0 &=&{\frac{8}{g_{A}^{2}}}\partial _{u}({\rm e}^{-2(\phi +\rho )}F_{uv})-
\sqrt{2N}\partial _{u}Y\ ,  \label{exxviii} \\
0 &=&{\frac{8}{g_{A}^{2}}}\partial _{v}({\rm e}^{-2(\phi +\rho )}F_{uv})-
\sqrt{2N}\partial _{v}Y\ .  \label{maxv}
\end{eqnarray}
The dilaton and matter equations of motion (\ref{eviii}) and (\ref{fermion})
are not modified. [As before, we consider the vacuum solution
(\ref{nomatter}
) to Eq. (\ref{fermion}).] The variables $Z$ and $Y$ satisfy the field
equations
\begin{equation}
0=-2Z_{,uv}+\sqrt{{\frac{N}{3}}}\rho _{,uv}\ ,  \label{exxixa}
\end{equation}
\begin{equation}
0=-2Y_{,uv}-\sqrt{2N}F_{uv}\ .  \label{exxixb}
\end{equation}

Motivated by the classical relation (\ref{Q}), we define $q$ by

\begin{equation}
\lambda q\equiv (\sqrt{2}/g_{A})~{\rm e}^{-2(\phi +\rho )}F_{uv}\,.
\label{sq}
\end{equation}
Note that $q$ is no longer constant: It evolves due to the semiclassical
source terms in the modified Maxwell equations (\ref{exxviii}, \ref{maxv}).
These two equations can immediately be integrated:

\begin{equation}
q=q_{0}+(g_{A}/4\lambda )\sqrt{N}\,Y\,,  \label{e51}
\end{equation}
where $q_{0}$ is an arbitrary integration constant. The equation for $Z$ can
be integrated too:

\begin{equation}
Z=\sqrt{{\frac{N}{12}}}\rho +Z_{v}(v)+Z_{u}(u)\ ,  \label{e52}
\end{equation}
where $Z_{v}(v)$ and $Z_{u}(u)$ are arbitrary initial functions.

To simplify the equations, we define $K\equiv N/24,$ $g\equiv \sqrt{3\,}
g_{A} $ , and rescale $Y$ and $Z$ as

\begin{equation}
z=Z/\sqrt{2{K}}\ ,\,\,y=Y/\sqrt{2{K}}\,.  \label{e53}
\end{equation}
Equations (\ref{exxixb},\ref{e51},\ref{e52}) then become, respectively,
\begin{equation}
y_{,uv}=-\sqrt{6}F_{uv}\ ,  \label{e54}
\end{equation}
\begin{equation}
q=q_{0}+(Kg/\lambda )\,y\,,  \label{e55}
\end{equation}
and
\begin{equation}
z=\rho +z_{v}(v)+z_{u}(u)\ ,  \label{e56}
\end{equation}
where again $z_{v}(v)$ and $z_{u}(u)$ are arbitrary initial functions. The
semiclassical contributions to the stress-energy now read

\begin{eqnarray}
\hat{T}_{uu} &=&K\left( z_{,u}^{\,2}-2\rho
_{,u}z_{,u}+z_{,uu}+y_{,u}^{\,2}\right) \,,  \label{e58a} \\
\hat{T}_{vv} &=&K\left( z_{,v}^{\,2}-2\rho
_{,v}z_{,v}+z_{,vv}+y_{,v}^{\,2}\right) \,,  \label{e58b} \\
\hat{T}_{uv} &=&-K\,\rho _{,uv}\ .  \label{e58c}
\end{eqnarray}
Using Eq. (\ref{sq}), we can rewrite the field equation for $y$ in terms of $
q$:
\begin{equation}
y_{,uv}=-\lambda g~{\rm e}^{2(\phi +\rho )}q\ .  \label{e60}
\end{equation}
The closed system of semiclassical field equations is composed of Eq. (\ref
{eq-fi}) for the dilaton, the three Einstein equations (\ref{svv}-\ref
{exxvii}), as well as Eqs. (\ref{e55}-\ref{e60}).

Finally, we write the system of semiclassical field equations in the $R,S$
variables. The dilaton equation (\ref{eq-fi}) and the Einstein equation (\ref
{exxvii}) [with Eq. (\ref{e58c})] yield
\begin{equation}
R_{,uv}=\lambda ^{2}({q}^{2}/R^{2}-1)~{\rm e}^{S}-K\,\rho _{,uv}\ ,
\label{e61}
\end{equation}

\begin{equation}
\ S_{,uv}=-(2\lambda ^{2}{q}^{2}/R^{3})~{\rm e}^{S}+K\,\rho _{,uv}/R\ ,
\label{e62}
\end{equation}
where $\rho =(S-\log R)/2$. The semiclassically-corrected constraint
equations take the form
\begin{eqnarray}
R_{,uu}-R_{,u}S_{,u}+\hat{T}_{uu} &=&0~,  \label{Ruu1} \\
R_{,vv}-R_{,v}S_{,v}+\hat{T}_{vv} &=&0~.  \label{Rvv1}
\end{eqnarray}
Again, this system is supplemented by Eqs. (\ref{e55} - \ref{e60}). Equation
(\ref{e60}) can be re-expressed using the $R,S$ variables as

\begin{equation}
y_{,uv}=-\lambda g\,({\rm e}^{S}/R^{2})\,q\ .  \label{e60a}
\end{equation}

\subsection{Semiclassical fluxes outside the black hole}

We turn now to analyze the evolution of the quantum variables $Y$ and $Z$,
in order to obtain the semiclassical fluxes. To that end we use the
adiabatic approximation. Namely, we view $Y$ and $Z$ as test fields living
on the background described by the static classical solution (with fixed $m$
and $q$).

We first calculate the semiclassical electric currents outside the BH. Using
Eq. (\ref{external}), we rewrite Eq. (\ref{e60a}) as

\begin{equation}
y_{,uv}=-\lambda gqH/R^{2}\ .  \label{e63}
\end{equation}
We now integrate this equation with respect to $v$, recalling $dv=dx=dR/H$:

\begin{equation}
y_{,u}=-\lambda gq\int (H/R^{2})dv=-\lambda gq\int R^{-2}dR=\lambda
gq/R+J_{u}(u)\ .  \label{e64}
\end{equation}
Similarly, we find for $y_{v}$ (recalling $du=-dx$):

\begin{equation}
y_{,v}=-\lambda gq\int (H/R^{2})du=\lambda gq\int R^{-2}dR=-\lambda
gq/R+J_{v}(v)\ .  \label{e65}
\end{equation}
The integration constants, i.e. the functions $J_{u}(u)$ and $J_{v}(v)$, are
to be determined from the initial conditions. Since we assume no ingoing
current is coming from past null infinity, we must set $J_{v}(v)=0$. Also,
regularity at the EH requires that $y_{,u}$ vanishes there, which implies $
J_{u}(u)=-\lambda gq/R_{+}$. Therefore,

\begin{equation}
y_{,u}=\lambda gq(1/R-1/R_{+})\quad ,\quad y_{,v}=-\lambda gq/R\,,
\label{e66}
\end{equation}
and from Eq. (\ref{e55}) we obtain

\begin{equation}
q_{,u}=Kg\,^{2}q(1/R-1/R_{+})\quad ,\quad q_{,v}=-Kg^{2}q/R\,.
\label{e66a}
\end{equation}

The fluxes $\hat{T}_{vv}$ and $\hat{T}_{uu}$ , Eqs.
(\ref{e58a},\ref{e58b}),
can be expressed explicitly by means of Eq. (\ref{e56}):

\begin{eqnarray}
\hat{T}_{uu} &=&K[(\rho _{,uu}-\rho _{,u}^{\,2})+y_{,u}^{\,2}\,+\hat{z}
_{u}(u)]\,,  \label{e67} \\
\hat{T}_{vv} &=&K[(\rho _{,vv}-\rho _{,v}^{\,2})+y_{,v}^{\,2}\,+\hat{z}
_{v}(v)]\,,  \label{e68}
\end{eqnarray}
where $\hat{z}_{v}(v)\equiv z_{v,vv}+z_{v,v}^{\,2}$ and $\ \hat{z}
_{u}(u)\equiv z_{u,uu}+z_{u,u}^{\,2}$. Substituting Eq. (\ref{e66}), we find
\begin{eqnarray}
\hat{T}_{uu} &=&K[(\rho _{,uu}-\rho _{,u}^{\,2})+\lambda
^{2}g^{2}q^{2}(1/R-1/R_{+})^{2}\,+\hat{z}_{u}(u)]\,, \\
\hat{T}_{vv} &=&K[(\rho _{,vv}-\rho _{,v}^{\,2})+\lambda
^{2}g^{2}q^{2}/R^{2}\,+\hat{z}_{v}(v)]\,.
\end{eqnarray}
The functions $\hat{z}_{v}(v)$ and $\hat{z}_{u}(u)$ are to be chosen such
that no influx is coming from past null infinity, and the outflow is regular
at $R=R_{+}$ , that is,
\begin{equation}
\hat{T}_{vv}(R=\infty )=0\quad ,\quad \hat{T}_{uu}(R=R_{+})=0\,.
\label{e69}
\end{equation}
In the static classical background we have $\rho =(1/2)\log (H/R)$, so
\begin{equation}
-\rho _{,u}=\rho _{,v}=\rho _{,x}=H\rho _{,R}=(R/2)(H/R)_{,R}  \label{e70}
\end{equation}
and
\begin{equation}
\rho _{,uu}=\rho _{,vv}=\rho _{,xx}=H[(R/2)(H/R)_{,R}]_{,R}  \label{e71}
\end{equation}
Note that $\rho _{,uu}$ and $\rho _{,vv}$ (as well as $\rho _{,uv}$) vanish
both at $R_{+}$ and at $R=\infty $. On the other hand, $\rho _{,u}$ and $
\rho _{,v}$ vanish at $R=\infty $, but at $R=R_{+}$ they get a finite value,
\begin{equation}
-\rho _{,u}=\rho _{,v}=\kappa _{+}\,,\quad \quad (R=R_{+})  \label{e72}
\end{equation}
In order for $\hat{T}_{vv}$ to vanish at $R=\infty $, we choose $\hat{z}
_{v}=0$ and obtain
\begin{equation}
\hat{T}_{vv}=K[(\rho _{,vv}-\rho _{,v}^{\,2})+(\lambda gq/R)^{2}\,]\,.
\label{e74}
\end{equation}
Also, the demand that $\hat{T}_{uu}$ vanishes at $R=R_{+}$ yields
$\hat{z}
_{u}=\kappa _{+}^{2}$ , namely,
\begin{equation}
\hat{T}_{uu}=K\left[ \kappa _{+}^{2}+(\rho _{,uu}-\rho
_{,u}^{\,2})+[\lambda
gq(1/R-1/R_{+})]^{2}\right] \,.  \label{e75}
\end{equation}

\subsection{Semiclassical fluxes inside the black hole}

Before we discuss semiclassical effects inside the BH, we need to extend our
classical solution for the static black-hole background to the internal
region. Clearly, Eq. (\ref{external}) as it stands is not valid at $
R_{-}<R<R_{+}$ , where $H$ is negative. The internal solution in double
null, Eddington-like coordinates is obtained form Eq. (\ref{external}) by
minor changes of sign. The main difference is that, inside the BH the
variable $x$ (the only variable on which the solution depends) is {\it \
temporal} rather than spatial, namely, $x=v+u$. The internal solution takes
the form

\begin{equation}
\ {\rm e}^{S}=-H(R)~,~R_{,x}=H(R)~,~x=v+u\qquad (R_{-}<R<R_{+})\,.
\end{equation}
Correspondingly, the metric function $g_{uv}$ is given by

\[
-2g_{uv}=e^{2\rho }=-H/R\,.
\]

We can now repeat the calculations of the semiclassical fluxes. The initial
conditions for the outgoing fluxes at $v\rightarrow -\infty $ are the same
as in the external problem: Both the energy and charge outfluxes must
vanish
at $R=R_{+}$, in order to achieve regularity at the EH. The initial
conditions for the {\it ingoing} fluxes at $u\rightarrow -\infty $ (the EH)
are dictated by continuity: At the EH, both $q_{,v}$ and $\hat{T}_{vv}$ must
continuously match the corresponding quantities in the external region $
R>R_{+}$ (recall that the coordinate $v$ is regular at the EH).

The calculation now proceeds in a way completely analogous to the external
semiclassical calculations of the previous subsection, except for a few
changes of sign. For example, when calculating the charge fluxes, one must
recall that ${\rm e}^{S}=-H$ and $du=dx$, and as a consequence $q_{,u}$
changes sign (but not $q_{,v}$):

\begin{equation}
q_{,u}=-Kg\,^{2}q(1/R-1/R_{+})\quad ,\quad q_{,v}=-Kg^{2}q/R\qquad
(R_{-}<R<R_{+})\,.
\end{equation}
The energy-momentum fluxes are given by Eqs. (\ref{e74}) and (\ref{e75})
without any change (recall, though, that now $\rho $ is given by $e^{2\rho
}=-H/R$).

\section{Evaporation of the black hole}

\subsection{Evolution of $m$ and $q$}

In order to calculate the rate of change of $m$ and $q$, as measured by an
observer at future null infinity (FNI), we need to evaluate the outgoing
fluxes at the limit $R\rightarrow \infty $. For brevity we denote the $u$
-derivatives of $m$ and $q$ at FNI by an overdot. At this limit Eq. (\ref
{e66a}) reads
\begin{equation}
\dot{q}=-Kg^{2}\,q/R_{+}\,.  \label{e82}
\end{equation}
Taking the large-$R$ limit in Eq. (\ref{e75}), we find

\begin{equation}
\hat{T}_{uu}=K[\kappa _{+}^{2}+(\lambda gq/R_{+})^{2}]=K\lambda
^{2}[\lambda
^{2}(m^{2}-q^{2})+g^{2}q^{2}]/R_{+}^{2}\quad \quad \rm{(FNI)}\,.
\label{Tuu}
\end{equation}
The relation between $\dot{m}$ and $\hat{T}_{uu}$ is most easily expressed
in terms of the BH's physical mass $M=2\lambda m$ and the
asymptotically-flat null coordinate $\tilde{u}=\lambda u$ (see section II):
The change in the Bondi mass $M$ is simply (minus) the integral of
$\hat{T}_{
\tilde{u}\tilde{u}}$ with respect to $\tilde{u}$ along FNI; that is,
\[
\partial M/\partial \tilde{u}=-\hat{T}_{\tilde{u}\tilde{u}}=-\lambda ^{-2}
\hat{T}_{uu}\,\quad \quad \rm{(FNI)}\,.
\]
We find that

\begin{equation}
\dot{m}=-\hat{T}_{uu}/2\lambda ^{2}\quad \quad \rm{(FNI)}\,.
\label{mdot}
\end{equation}
Alternatively, we can derive this relation using the Vaidya-like solution
constructed in Appendix A. To that end, we recall that the ingoing fluxes of
both energy and charge, as well as the semiclassical correction term
$\hat{T}
_{uv}$ , vanish at FNI (where $R\rightarrow \infty $) -- cf. Eqs. (\ref{e66},
\ref{e74}). The only semiclassical correction terms which survive at FNI are
the Hawking energy outflux $\hat{T}_{uu}$ and charge outflux $q_{,u}$. We
can therefore represent the solution near FNI by the (charged) outgoing
Vaidya-like solution. This exact solution provides the desired relation
between $\dot{m}$ and the energy outflux at FNI. Identifying $\hat{T}_{uu}$
with $T_{uu}^{(flux)}$ in Eq. (\ref{vm}) below, we recover the relation (\ref
{mdot}).

Substituting the above expression for $\hat{T}_{uu}($FNI$)$ in Eq. (\ref
{mdot}), we find

\begin{equation}
\dot{m}=-K[\kappa _{+}^{2}/\lambda ^{2}+(gq/R_{+})^{2}]/2=-K[\lambda
^{2}(m^{2}-q^{2})+g^{2}q^{2}]/2R_{+}^{2}\,.  \label{e81}
\end{equation}

To verify the consistency of the above results for $\dot{m}$ and $\dot{q}$
(and, more generally, for the fluxes of energy-momentum and charge), we
can
calculate the rate of change of \thinspace $R_{+}$ in two different ways.
First, since the evaporation is very slow (corresponding to a large,
macroscopic, BH), we can view the geometry as quasi-static. At each
''moment'' $u$, $R_{+}$ (as viewed by a distant observer) can be estimated
by the momentary values of $m$ and $q$, via the standard, static-solution
relation $R_{+}=m+(m^{2}-q^{2})^{1/2}$. Alternatively, we can apply the
constraint equation $R_{,vv}-R_{,v}S_{,v}+\hat{T}_{vv}=0$ to the null
generators of the EH, and in this way to analyze the rate of contraction of
$
R_{+}$. Since the evolution is slow (and hence, on time scales short
compared to the BH's evaporation time, the geometry only depends on $x=v-
u$
to the leading order), $\partial R_{+}/\partial v$ at the EH must coincide
with $\dot{R}_{+}\equiv \partial R_{+}(m,q)/\partial u$ at FNI, i.e. with $
\dot{m}\,R_{+,m}+\dot{q}\,R_{+,q}$. In Appendix B we calculate these two
quantities and show they are indeed the same:

\begin{equation}
R_{+,v}(\rm{EH})=\dot{R}_{+}=-K\lambda ^{2}[\lambda
^{2}(m^{2}-q^{2})-g^{2}q^{2}\,]/(2\kappa _{+}R_{+}^{2})\,.  \label{rdvrdu}
\end{equation}

Finally, let us compare our result (\ref{Tuu}) for the Hawking outflux at
FNI to the standard result obtained by CGHS (in the uncharged case). Taking
the limit $q=0$ in Eq. (\ref{Tuu}), one finds

\begin{equation}
\hat{T}_{uu}=K\lambda ^{4}m^{2}/R_{+}^{2}=K\lambda ^{4}/4\quad \quad
\quad (
\rm{FNI}\,,\,q=0)\,.  \label{q0}
\end{equation}
Transforming this result to the asymptotically-flat $\tilde{u},\tilde{v}$
coordinates defined in section II (which is also the gauge used by CGHS), we
find

\begin{equation}
\hat{T}_{\tilde{u}\tilde{u}}=K\lambda ^{2}/4=N\lambda ^{2}/96\quad \quad
\quad (\rm{FNI}\,,\,q=0)\,.  \label{e80}
\end{equation}
This is just one half of the Hawking outflux in the CGHS model. This
difference is because the quantum matter field used in the present model is
fermionic, whereas that used in the CGHS model is bosonic \cite{oda}.

\subsection{Evolution of the charge-to-mass ratio}

Equations (\ref{e82}) and (\ref{e81}) form a closed autonomous system,
which
allows us to analyze the evolution of the charge-to-mass ratio upon
evaporation. One finds
\begin{equation}
dm/dq=(R_{+}/2q)\left[ (\kappa _{+}/\lambda
g)^{2}+q^{2}/R_{+}^{2}\right] \,.
\label{e83}
\end{equation}
Since the right-hand side only depends on $q$ and $m$ through $q/m$, this
equation admits solutions of the form
\begin{equation}
m=cq\,,  \label{e84}
\end{equation}
where $c$ is a positive constant [to be determined from an algebraic
equation based on Eq.(\ref{e83}), as we shortly show]. We shall refer to a
solution of this form as the{\it \ linear solution}. To analyze this
solution, we rewrite Eq. (\ref{e83}) as

\begin{equation}
dm/dq=(R_{+}/2q)\left[ (\kappa _{+}/\lambda g)^{2}-1\right] +m/q\,.
\label{e85}
\end{equation}
This form makes it obvious that for any $\,0<g<\lambda /2$, there exists a
linear solution of the form (\ref{e84}), with $c$ defined by the algebraic
equation $\kappa _{+}(m=cq)=\lambda g$, i.e.
\begin{equation}
g/\lambda \,=1-\left( 1+\sqrt{1-c^{-2}}\right) ^{-1}\,
\end{equation}
[cf. Eq. (\ref{e73a})]. Explicitly we find
\begin{equation}
c=\left( 1-[(1-g/\lambda )^{-1}\,-1]^{2}\right) ^{-1/2}\,\quad \quad
(0<g<\lambda /2)\;.  \label{e86}
\end{equation}

Next we analyze the stability of the linear solution (\ref{e84}, \ref{e86}).
To that end, we define $\delta \equiv m/q-c$, and write the evolution
equation for $\delta $ in the form
\begin{equation}
\frac{d\delta }{d\ln q}=\frac{dm}{dq}-\frac{m}{q}=(R_{+}/2q)\left[ (\kappa
_{+}/\lambda g)^{2}-1\right] \equiv \Gamma (\delta )\,.  \label{e87}
\end{equation}
Note that $\kappa _{+}$ is an increasing function of $\delta $ [cf. Eq. (\ref
{e73a})]. Since $\lambda g=\kappa _{+}(m/q=c)=\kappa _{+}(\delta =0)$,
the
term in squared brackets is an increasing function of $\delta $ which
vanishes for $\delta =0$. Therefore, $\Gamma $ has the same sign as
$\delta $
, which means that $|\delta |$ is an increasing function of $q$. This
implies that upon evaporation ($q$ decreases), $|\delta |$ decreases.
Namely, the linear solution (\ref{e84}, \ref{e86}) is stable. Moreover,
since the only zero of $\Gamma $ is at $\delta =0$, the linear solution is
in fact a global atractor for any $0<g<\lambda /2$, provided that initially $
q>0$. [For small $\delta $, we can linearize $\Gamma $ by $\Gamma \cong
\beta \delta $, with some constant $\beta =\beta (g)>0$. We then find that
$
\delta \varpropto q^{\beta }\varpropto m^{\beta }$.]

The dynamical system (\ref{e82},\ref{e81}) has another, trivial, solution:
\begin{equation}
q/m=0\,.
\end{equation}
For $0<g<\lambda /2$, this solution must be unstable: As we have just
found,
the linear solution (\ref{e84},\ref{e86}) is a global atractor in this range
for any initial $q>0$. However, for $g>\lambda /2$, for which the above
linear solution does not exist ($c$ is not real), the trivial solution $q=0$
becomes stable. To verify this, we define in this case $\delta \equiv m/q>0$
, and analyze Eq. (\ref{e87}). Since now $\kappa _{+}/\lambda g<1$ (for any
$
q/m$), $\Gamma (\delta )$ is always negative, meaning that upon
evaporation $
\delta $ increases monotonically. Furthermore, since $R_{+}/m\geq 1$, the
quantity $d\ln \delta /d\ln q=\Gamma /\delta $ is bounded above by the
strictly negative number $\gamma /2$, where $\gamma \equiv \lambda
^{2}/4g^{2}-1<0$. This means that upon evaporation ($\ln q\rightarrow
-\infty $), $\delta $ gets unboundedly large positive values. Once $\delta $
becomes large, we can use the linear approximation $\Gamma /\delta \cong
\gamma $ (obtained by approximating $R_{+}\cong 2m$ and $\kappa
_{+}\cong
\lambda ^{2}/2$), which yields $\delta \varpropto q^{\gamma }$. This
implies
$m\varpropto q^{\lambda ^{2}/4g^{2}}$, namely,
\begin{equation}
q/m\varpropto m^{4g^{2}/\lambda ^{2}-1}\quad \quad (g>\lambda /2)\,.
\label{smallq}
\end{equation}
It should be pointed out that this linear analysis of solutions with $q/m<<1$
, and particularly the result (\ref{smallq}), applies to {\it any} $g$. It
indicates the stability of the solution $q/m=0$ in the range $g>\lambda /2
$
, and its instability in the range $g<\lambda /2$. Thus, for $g>\lambda /2$,
Eq. (\ref{smallq}) is realized as the late-time, stable, asymptotic
behavior. For $g<\lambda /2$, however, even if initially $q/m<<1$, upon
evaporation it grows according to Eq. (\ref{smallq}) until the linear
approximation breaks (provided, of course, that initially $q>0$).
Subsequently $q/m$ converges to a nonzero value $c^{-1}$, as was
discussed
above.

We conclude that for $g<\lambda /2$, the charge-to mass ratio converges
to a
nonzero value,

\begin{equation}
\frac{q}{m}\rightarrow \sqrt{1-[(1-g/\lambda )^{-1}\,-1]^{2}}\,\quad \quad
(0<g<\lambda /2)\;.  \label{smallg}
\end{equation}
This value is independent of the initial values of $q$ and $m$ (though it
only holds if $q$ is initially nonzero). However, for $g>\lambda /2$, the
charge-to mass ratio decreases as a power of $m$, and eventually
approaches
zero, as described in Eq. (\ref{smallq}).

\section{Summary}

We presented here a dilatonic two-dimensional model of a charged black hole.
On the classical level, our model yields static charged BHs, characterized
by the two parameters $m$ and $q$ (representing the BH's mass and
charge).
These static BHs admit an inner horizon instead of a spacelike singularity.
Then we studied the semiclassical effects (on the background of the above
classical, static, BH solution), using the method developed in Ref. \cite{NO}
. We derived explicit expressions for the fluxes of charge and
energy-momentum as a function of the ''radius'' $R$, both outside and inside
the BH.

By analyzing the outflux of energy-momentum and charge at future null
infinity (and also the influx at the EH), we calculated the evaporation rate
of both $m$ and $q$, as a function of these two quantities. This yields a
system of two coupled first-order differential equations, i.e. $\dot{q}(m,q)$
and $\dot{m}(m,q)$ [Eqs. (\ref{e82}) and (\ref{e81}), respectively]. We then
analyzed the evolution of the ratio $q/m$ upon evaporation. Depending on the
value of the electromagnetic coupling constant $g$ (recall $g\equiv
\sqrt{3}
\,g_{A}$), there are two different regimes: For $g>\lambda /2$, upon
evaporation $q/m$ decays to zero as described in Eq. (\ref{smallq}) above.
On the other hand, for $g<\lambda /2$, the charge-to-mass ratio
approaches a
non-zero constant given in Eq. (\ref{smallg}). This constant depends on $g$
but not on the initial values of $m$ and $q$ (provided that $q$ is initially
nonvanishing). Note that this final charge-to-mass ratio approaches
extremality ($q/m=1$) at the limit $g\rightarrow 0$, and $q/m\rightarrow
0$
at the limit $g\rightarrow \lambda /2$.

As was explained in the Introduction, no attempt was made here to
investigated the final state of evaporation. The analysis throughout this
paper was restricted to the macroscopic phase, i.e. to the stage where the
mass is sufficiently large. This restriction is necessary for the validity
of the adiabatic approximation: This approximation assumes that in
evaluating the semiclassical fluxes (more specifically, when solving the
field equations for the quantum variables $Y$ and $Z$), $m$ and $q$ may be
regarded as fixed parameters (and the background geometry may be
approximated by the corresponding static BH solution). Clearly, this
approximation is only valid as long as the change in $m$ during a dynamical
time scale is much smaller than $m$ itself. The dynamical time scale
(expressed in terms of $u$ and/or $v$) is of order $1/\kappa _{+}$ , which
is typically of order $\sim \lambda ^{-2}$. The mass evaporation rate
$\dot{m
}$ is of order $K\lambda ^{2}$ (recall $K\equiv N/24$). Thus, the
macroscopic phase -- the domain of validity of the adiabatic approximation
-- is given by
\begin{equation}
m\gg K\,.  \label{mmac}
\end{equation}
In this domain the dilaton field $\phi $ outside the BH satisfies
\[
e^{-2\phi }\gg K\,.
\]
(This also holds inside the BH, in the region $R>R_{-}$ -- provided that $
q/m $ is not too small.) This is known to be the ''weak-coupling'' domain in
large-$N$ dilaton gravity. It should also be pointed out that the curvature
singularity found in Ref. \cite{russo} (in the uncharged case) occurs at $
e^{-2\phi }=2K$, which does not occur in the macroscopic domain considered
here.

\section*{Acknowledgment}

I would like to thank Eanna Flanagan, Joshua Feinberg, Valeri Frolov, Andrei 
Zelnikov, Don Page, Shin'ichi Nojiri, and Ichiro Oda for interesting discussions 
and helpful advise. 
This research was supported in part by the United States-Israel Binational 
Science Foundation.

\appendix

\section{Vaidya-like solutions}

In this Appendix we describe the construction of the Vaidya-like solution in
double-null coordinates. For concreteness we shall consider here the
outgoing solution, which is useful for describing the geometry of an
evaporating BH (in the weak-field region), but the ingoing solution can be
constructed in a completely analogous manner.

Consider first the case of an uncharged null fluid (i.e. $q$ is a fixed
parameter). The solution is uniquely determined by the function $m(u)$. In
analogy with the construction of the static vacuum-like solutions in section
II, we define

\begin{equation}
H(R,u)\equiv \lambda ^{2}\left[ R-2m(u)+q^{2}/R\right] \,.  \label{vh}
\end{equation}
The function $R(u,v)$ is now determined by the ordinary differential equation

\begin{equation}
R_{,u}=-H(R,u)\,.  \label{vR}
\end{equation}
This equation is to be integrated along the lines of constant $v$ (with each
ingoing null line having its own ''initial value'' for $R$), and this
integration produces the function $R(u,v)$.\footnote{
This construction fixes the gauge for the outgoing coordinate $u$, but
leaves the gauge for the ingoing coordinate $v$ uncpecified. In this
construction $v$ enters as a parameter which parametrizes the one-
parameter
set of solutions to the differential equation (\ref{vR}) for $R$. (For
example, one can take $v$ to be the value of $R$ on some ''initial''
outgoing ray $u=u_{0}$.)} Then, $S(u,v)$ is given by
\begin{equation}
e^{S}=R_{,v}\,.  \label{vS}
\end{equation}
Differentiating Eq. (\ref{vR}) with respect to $v,$ one recovers the field
equation (\ref{universal}) for $R$, i.e. $R_{,uv}=Fe^{S}$. Next,
differentiating Eq. (\ref{vS}) with respect to $u$, one obtains $S_{,u}=F$,
and a second differentiation with respect to $v$ now yields the field
equation for $S$.

To discuss the null-fluids content of the Vaidya-like solutions (or, more
generally, the radiative solutions), it is useful to re-write the constraint
equations as

\begin{eqnarray}
\ 0 &=&T_{uu}=R_{,uu}-R_{,u}S_{,u}+T_{uu}^{(flux)}~, \\
0 &=&T_{vv}=R_{,vv}-R_{,v}S_{,v}+T_{vv}^{(flux)}~.
\end{eqnarray}
The components $T_{uu}^{(flux)}$ and $T_{vv}^{(flux)}$ then describe the
energy-momentum carried by the outgoing and ingoing null fluids,
respectively. (The vacuum-like solutions then correspond to $
T_{uu}^{(flux)}=T_{vv}^{(flux)}=0$.) Differentiating Eq. (\ref{vR}) with
respect to $v$ yields $R_{,v}S_{,v}=R_{,vv}$. Namely, the outgoing
Vaidya-like solution is characterized by a vanishing influx:
\begin{equation}
T_{vv}^{(flux)}=0\,.  \label{vTvv}
\end{equation}
To express the flux in the $u$ direction in terms of $m(u)$, we notice that
the above result $S_{,u}=F$ together with Eq. (\ref{vR}) yield $
R_{,u}S_{,u}=-FH\,$ and $R_{,uu}=-FH\,-\partial H/\partial u$ [where $
\partial H/\partial u\equiv (\partial H/\partial m)(\partial m/\partial u)$
], and hence
\begin{equation}
T_{uu}^{(flux)}=R_{,u}S_{,u}-R_{,uu}=\partial H/\partial u=-2\lambda
^{2}m_{,u}\,.  \label{vTuu}
\end{equation}
Note that $T_{uu}^{(flux)}$ is independent of $v$ -- this is an important
feature of the Vaidya-like solutions [valid for any $F(R)$].

Strictly speaking, the field equations in the form (\ref{universal},\ref
{F(R)}) assume that $q$ is a constant. One can, however, immediately
generalize it by allowing $q$ to be a function of $u$ and $v$. When
considering the (outgoing) Vaidya-like solutions, it is most natural to
assume that $q$ (like $m)$ depends on $u$ only. Physically, this would
correspond to a model with an outflow of charged null fluid. This
generalization is important because in our model the Hawking outflux is
indeed charged.

The generalization of the Vaidya-like solution to the charged null fluid
case is straightforward. One simply replaces Eq. (\ref{vh}) by

\begin{equation}
H(R,u)\equiv \lambda ^{2}\left[ R-2m(u)+q(u)^{2}/R\right] \,.  \label{vhq}
\end{equation}
The rest of the above construction, Eqs. (\ref{vR}) and (\ref{vS}), are
unchanged. The solution is uniquely determined by the two functions $m(u)$
and $q(u)$, which describe the outflux of mass and charge, respectively. One
can verify that the evolution equations (\ref{eqR},\ref{eqS}) are satisfied,
as well as Eq. (\ref{vTvv}). However, the energy-momentum content of the
outgoing flux is now modified:

\begin{equation}
T_{uu}^{(flux)}=R_{,u}S_{,u}-R_{,uu}=\partial H/\partial u=-2\lambda
^{2}\left( m_{,u}-q\,q_{,u}/R\right) \,.  \label{vTvvq}
\end{equation}
Note that $T_{uu}^{(flux)}$ is no longer constant along lines of constant $u$
. This has a simple physical interpretation: The Lorentz force acting on the
charged outflux does a work on it, and changes its energy-momentum
content.
(This situation is well known in the context of the four-dimensional,
spherically symmetric, charged Vaidya solution\cite{bonnor}; see the
discussion in \cite{sullivan} and\cite{ori}.) Note, however, that at FNI the
term$\,q_{,u}/R$ vanishes, and one again obtains
\begin{equation}
T_{uu}^{(flux)}=-2\lambda ^{2}m_{,u}\quad \quad \rm{(FNI)}\,.  \label{vm}
\end{equation}
We can verify this result by relating $T_{uu}^{(flux)}$ to the rate of
change of the Bondi mass $M$. In terms of the asymptotically-flat null
coordinate $\tilde{u}$, these two quantities are related by
\[
\partial M/\partial \tilde{u}=-\hat{T}_{\tilde{u}\tilde{u}}\,.
\]
Substituting $M=2\lambda m$, $\tilde{u}=\lambda u$, and
$\hat{T}_{\tilde{u}
\tilde{u}}=\lambda ^{-2}\hat{T}_{uu}$, we find $2m_{,u}=\partial M/\partial
\tilde{u}=-\lambda ^{-2}\hat{T}_{uu}$ , which conforms with Eq. (\ref{vm}).

The Vaidya-like solution may be interpreted as a slowly varying,
quasi-static solution, which is described by the vacuum-like solution (\ref
{external}) -- except that the BH's mass and charge are slowly evaporating.
This interpretation is meaningful as long as the evaporating BH is
macroscopic (i.e. the evaporation time scale is much larger than the
dynamical time scale). Note that in this quasi-static limit the coordinate $
u $ used in the above construction of the Vaidya-like solution coincides
with the Eddington-like coordinate $u$ of the static, vacuum-like, solution (
\ref{external}). That the gauges of these two solutions agree can be seen,
for example, by recognizing that $R_{,u}=-H$ in both solutions.

\section{Rate of contraction of the event horizon}

In this Appendix we shall calculate the contraction rate of \thinspace $
R_{+} $ in two different ways, as outlined in section IV. The equality of
the two results may serve as a consistency test for the expressions derived
above for $\dot{m}$ and $\dot{q}$.

First, since the evaporation is slow, we may assume that at each moment
$u$,
$R_{+}$ is given by
\[
R_{+}(u)=m(u)+[m(u)^{2}-q(u)^{2}]^{1/2}\,.
\]
Taking $u$-derivatives of all quantities at FNI, one finds

\begin{equation}
\dot{R}_{+}=\dot{m}\left[ 1+m/(m^{2}-q^{2})^{1/2}\right] -\dot{q}
q/(m^{2}-q^{2})^{1/2}\,.  \label{Rdot}
\end{equation}
Noting that
\[
1+m/(m^{2}-q^{2})^{1/2}=R_{+}/(m^{2}-q^{2})^{1/2}=\lambda
^{2}/\kappa _{+}
\]
[cf. Eq. (\ref{e73b})], we can re-write Eq. (\ref{Rdot}) as
\[
\kappa _{+}\dot{R}_{+}=\lambda ^{2}(\dot{m}-\dot{q}q/R_{+})\,.
\]
Substituting the above expressions for $\dot{m}$ and $\dot{q}$, one finds

\begin{equation}
\kappa _{+}\dot{R}_{+}=-K\lambda ^{2}[\lambda
^{2}(m^{2}-q^{2})-g^{2}q^{2}]/2R_{+}^{2}\,.  \label{Rdu}
\end{equation}

In the second way, we apply the constraint equation (\ref{Rvv1}) to the null
generators of the EH, and use it to analyze the rate of decrease of $R_{+}$
with $v$. Since the evolution is assumed to be slow, we can neglect the term
$R_{,vv}$ and write
\[
S_{,v}\,R_{+,v}=\hat{T}_{vv}\quad \quad (EH)\,.
\]
Using the background solution (\ref{external}) one can easily verify that at
the EH $S_{,v}=dH/dR=2\kappa _{+}$ , and Eq. (\ref{e74}) yields
\begin{equation}
\hat{T}_{vv}=-K[\kappa _{+}^{2}-(\lambda gq/R_{+})^{2}\,]\quad \quad
(EH)\,.
\end{equation}
Therefore,
\begin{equation}
\kappa _{+}\,R_{+,v}=\hat{T}_{vv}/2=-K\lambda ^{2}[\lambda
^{2}(m^{2}-q^{2})-g^{2}q^{2}\,]/2R_{+}^{2}\quad \quad (EH)\,.  \label{Rdv}
\end{equation}
Comparing Eqs. (\ref{Rdu}) and (\ref{Rdv}), we find that indeed $R_{+,v}$ at
the EH is exactly the same as $\dot{R}_{+}$ , as one should expect.

\end{document}